\documentclass[preprint,authoryear,12pt]{elsarticle}

 \usepackage{graphicx}
\usepackage{epsfig,lscape,rotating}

\usepackage{amssymb}

\usepackage[ps2pdf,%
a4paper=true,%
breaklinks=true,%
colorlinks=true,%
pdfauthor={M. A. Barstow et al.},%
pdftitle={The status and future of EUV astronomy}%
]{hyperref}

\journal{Advances in Space Research}

\begin{document}

\begin{frontmatter}



\title{The status and future of EUV astronomy}


\author{M. A. Barstow\corref{cor}}
\author {S. L. Casewell}
\address{Department of Physics and Astronomy, University of Leicester, University Road, Leicester, LE1 7RH, UK}
\cortext[cor]{Corresponding Author: mab@le.ac.uk}


\author{J. B. Holberg}
\address{Lunar and Planetary Laboratory, 1541 East University Boulevard, Sonett Space Sciences Building, University of Arizona, Tucson, AZ 85721, USA}

\author{M.P. Kowalski}
\address{Naval Research Laboratory,4555 Overlook Ave SW  Washington, DC 20375, USA}

\begin{abstract}
The Extreme Ultraviolet wavelength range was one of the final windows to
be opened up to astronomy. Nevertheless, it provides very important
diagnostic tools for a range of astronomical objects, although the opacity
of the interstellar medium restricts the majority of observations to
sources in our own galaxy. This review gives a historical overview of EUV
astronomy, describes current instrumental capabilities and examines the
prospects for future facilities on small and medium-class satellite
platforms.

\end{abstract}


\end{frontmatter}

\parindent=0.5 cm

\section{Introduction}
The Extreme Ultraviolet (EUV) nominally spans the wavelength range from 100 to
912 \AA, although for practical purposes the edges are often somewhat
indistinct as instrument band-passes may extend short-ward into the soft X-ray
or long-ward into the far ultraviolet (far-UV). The production of EUV photons
is primarily associated with the existence of hot, 10$^{5}$-10$^{7}$ K gas in the
Universe. Sources of EUV radiation can be divided into two main categories,
those where the emission arises from recombination of ions and electrons in a
hot, optically thin plasma, giving rise to emission line spectra, and objects
which are seen by thermal emission from an optically thick medium, resulting
in a strong continuum spectrum which may contain features arising from
transitions between different energy levels or ionisation stages of several
elements. Examples of the former category are single stars and binary systems
containing active coronae, hot O and B stars with winds, supernova remnants
and galaxy clusters. Hot white dwarfs, central stars of planetary nebulae
(CPN) and neutron stars are all possible continuum sources and belong to the
latter category. Cataclysmic variable binaries, where material is being transferred
from a normal main sequence star (usually a red dwarf) onto a white dwarf, may
well contain regions of both optically thin and optically thick
plasma. Perhaps the most unique contribution EUV observations can make to
astrophysics in general, is by providing access to the most important
spectroscopic features of helium - the He I and He II ground state continua
together with the He I and He II resonance lines. These are the best
diagnostics of helium, the second most cosmically abundant element. The
line series limits are at 504 \AA~ and 228 \AA~ for He I and He II,
respectively. 

\begin{figure}

\begin{center}
\includegraphics*[scale=1,angle=-0]{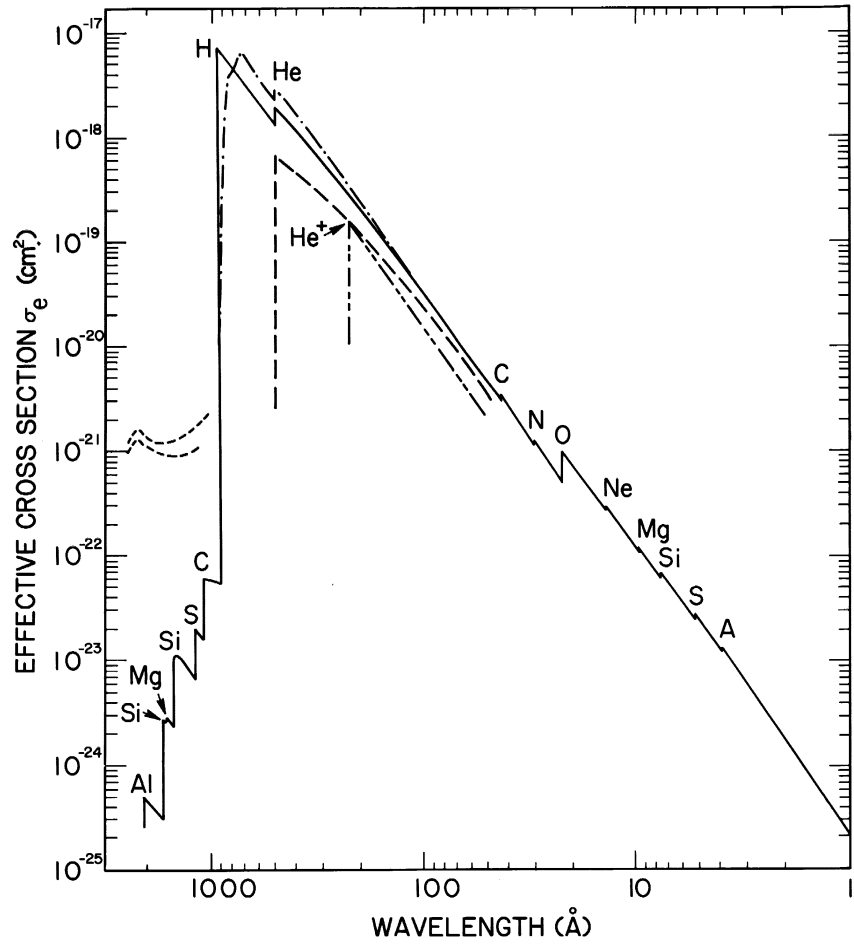}
\end{center}
\caption{\label{figure1} Effective cross-section of the interstellar medium.
------ gaseous component with
normal composition and temperature; --$\cdot $-- hydrogen, molecular
form; ---\ --- H II region about a B star; --\ --\ -\ - H II region
about an O star; - \ - \ - \ - \  dust
 (from \citealt{cruddace74}.)}
\end{figure}

Since most elements have outer electron binding energies in the
range 10-100 eV, corresponding closely to the energies of EUV photons, any EUV
radiation will interact strongly in any material through which it passes,
including the interstellar medium (ISM). Early measurements of the density of
interstellar gas indicated that it would be completely opaque to any EUV
radiation arising from outside the Solar System \citep{aller59}. This view
remained unchallenged for more than a decade, but increased understanding of
the ISM indicated a high degree of patchiness and significantly lower
densities in some directions than first assumed. Coupled with a more
sophisticated analysis of the absorption cross-sections across the EUV range
by (\citealt{cruddace74}, Figure \ref{figure1}), it became evident that EUV
sources should be detectable out to distances of a few hundred parsecs and at
the shortest wavelengths (bordering on the soft X-ray) some extragalactic
observations are possible. 

This review outlines the development of EUV
astronomy during the past 40 years, covering the early experiments on
sub-orbital missions, through the major satellite-borne survey missions, to
the development of spectroscopic instruments. Much of this history has been
comprehensively covered in \citet{barstow03}. Therefore, we summarise just the
main elements here, concentrating on the most recent developments and the
prospects for new missions, not included in that book.

\section{Sounding rockets and early missions}
With the realisation that galactic sources of EUV radiation may be detectable
from above the Earth's atmosphere, a number of investigators, mainly at the
Space Sciences Laboratory of the University of California, Berkeley, embarked
on a series of sounding rocket missions during the early 1970s. These
experiments did not use imaging instruments, although flux concentrators were
used, and had limited sensitivity. The sub-orbital nature of the programme
also restricted the observing time to just a few minutes. While several
thousand square degrees of the sky were surveyed, only a single source, most
likely associated with the cataclysmic variable VW Hyi, was detected \citep{henry76}.

An opportunity to carry out a more extended search for EUV sources arose in
1975 through the Soviet-American Apollo Soyuz mission (officially termed the
Apollo Soyuz Test Project, ASTP), a well publicised link up between US and
Soviet astronauts in orbit, as a demonstration of international good will. A
number of science instruments were allowed to piggyback on the Apollo capsule
for the 9 day duration of the mission. Two EUV instruments were included in
the programme, the Extreme Ultraviolet Telescope (EUVT, \citealt{bowyer77a})
and the Interstellar Helium Glow Experiment \citep{bowyer77b}. More than 30
potential EUV sources were observed, with exposure times up to 20
minutes, yielding detections of four prototypical sources: two hot white
dwarfs, HZ43 \citep{lampton76} and Feige 24 \citep{margon76}, the cataclysmic
variable SS Cyg \citep{margon78} and the nearby flare star Proxima Centauri
\citep{haisch77}. ASTP clearly demonstrated the feasibility of EUV astronomy
and that new missions with greater sensitivity could explore the wavelength
range with expectation of success.

\begin{figure}

\begin{center}
\includegraphics*[scale=1.0,angle=-0]{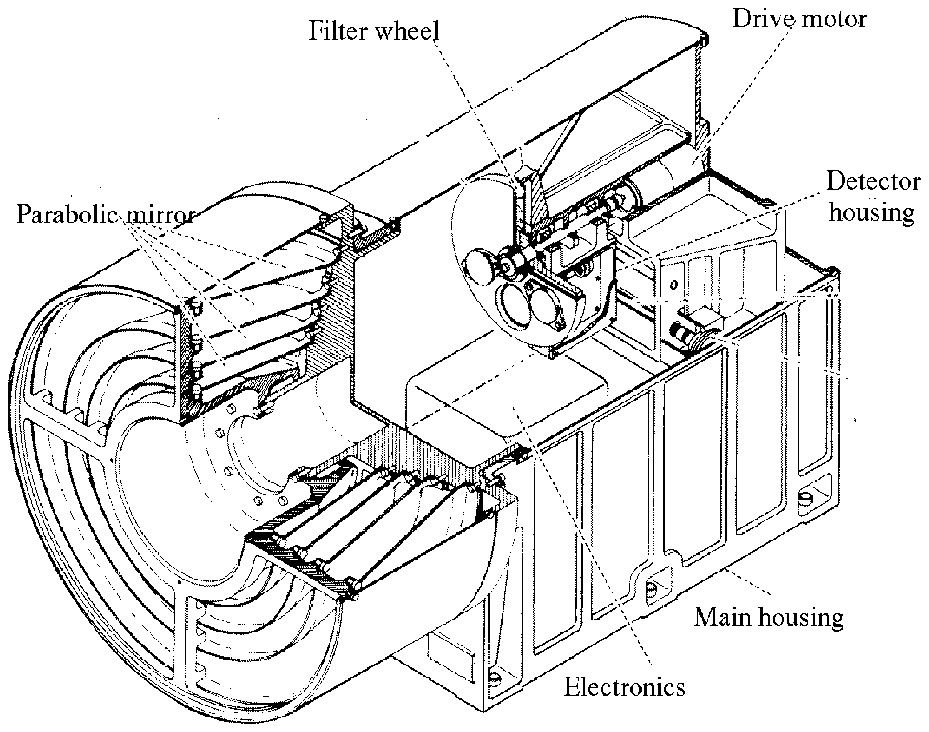}
\end{center}
\caption{\label{figure2} A schematic diagram of the  EUV  telescope  flown on board  the  Apollo-Soyuz mission  in 1975.}
\end{figure}

\section{Sky surveys}

While flux concentrators (such as ASTP) could collect photons over a
significantly large sky area, they were unable to discriminate between
background and source photons. The development of true imaging systems through
grazing incidence mirror technology in the late 1970s had an important impact
on the sensitivity of X-ray and EUV instrumentation, since the background
photons could be subtracted from the source image. Coinciding with these was
the growing need to find out just what sources of cosmic EUV radiation were
present to determine the scope and importance of EUV observations in
astrophysics. Therefore, in common with all new energy windows in astronomy,
all-sky surveys were planned for the EUV wavelength range. Two projects were
adopted, the NASA Extreme Ultraviolet Explorer (EUVE) and the UK Wide Field
Camera (WFC), to be flown as part of the German-led ROSAT X-ray sky survey
mission. 

\begin{figure}

\begin{center}
\includegraphics*[width=10cm,angle=-0]{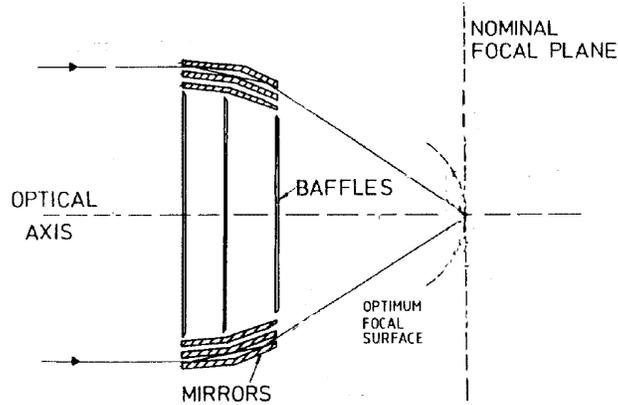}
\end{center}
\caption{\label{figure3}Schematic diagram of the optical systems of the ROSAT WFC.}
\end{figure}

These missions had similar objectives, to search the whole sky for EUV
sources, but slightly different technical approaches. The WFC
(\citealt{sims90}, Figure \ref{figure3}) was a single telescope with 3 nested
mirrors, designed to cover a wavelength range from 60 to 200 \AA, at the short
end of the EUV band, where the ISM is most transparent and most sources expect
to be visible. However, the instrument also had a capability for observing
longer wavelengths following the survey where the telescopes were pointed
directly at specific targets.

 \begin{figure}
\begin{center}
\includegraphics*[width=10cm,angle=-0]{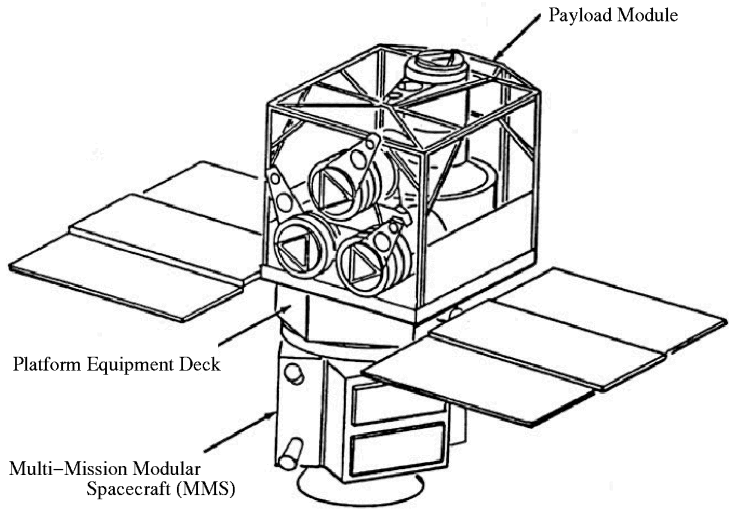}
\end{center}
\caption{\label{figure4}Schematic diagram of the EUVE payload,  showing  the  all-sky  survey scanners and  the deep  survey/spectrometer telescope  pointing at  right-angles to these.}
\end{figure} 
 \begin{figure}
\begin{center}
\includegraphics*[scale=0.9,angle=-0]{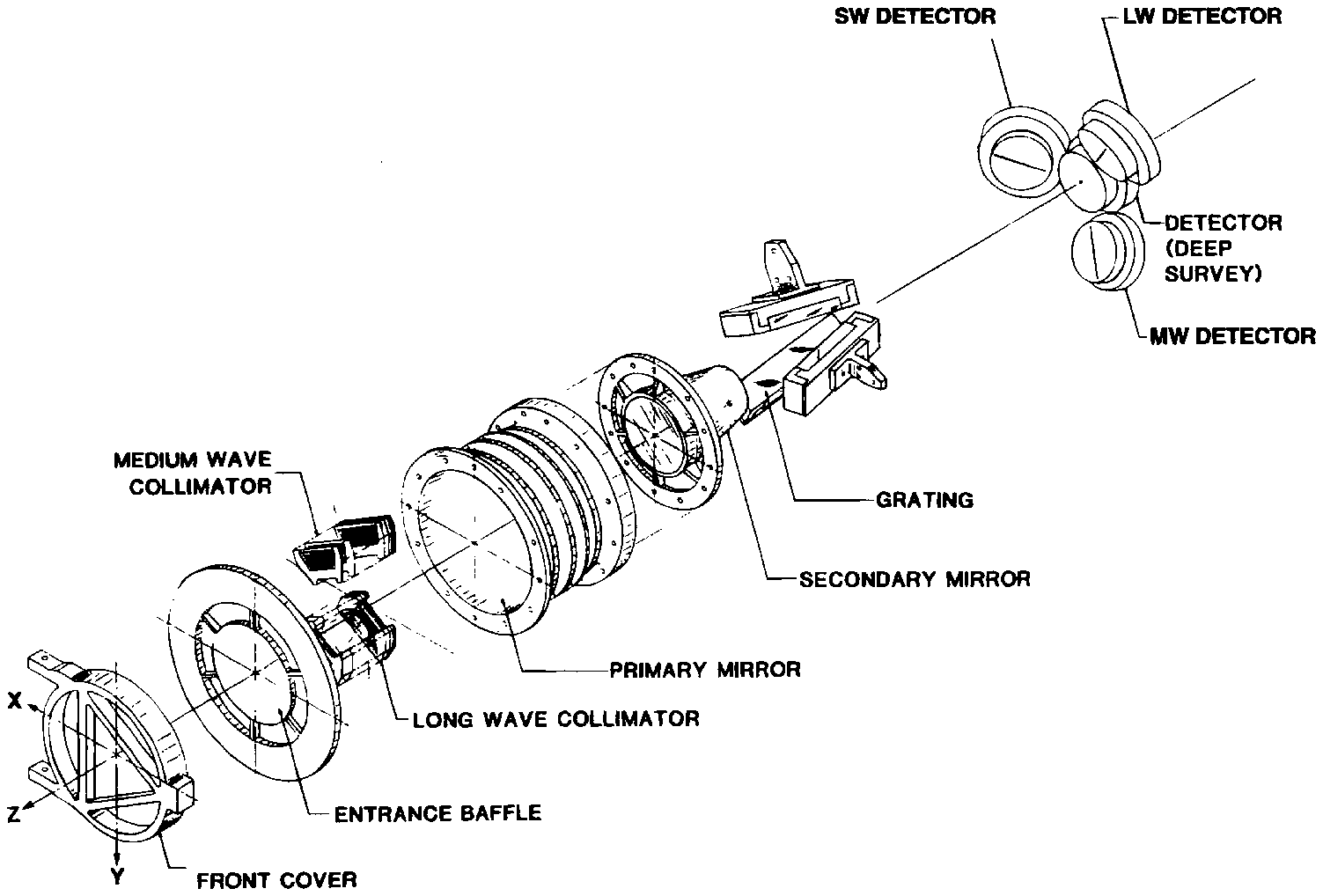}
\end{center}
\caption{\label{figure5}Exploded view of the EUVE Deep  Survey/Spectrometer telescope.  The axes drawn  at  the  front of the instrument represent the DSS coordinate system.}
\end{figure} 

EUVE (\citet{bowyer91a, bowyer91b}; Figure \ref{figure4}, \ref{figure5})
 had a complement of three single grazing incidence telescopes covering
 different wavelength ranges and providing all sky survey coverage extending
 to 800 \AA. In addition, EUVE carried a spectrometer for pointed observations
 following the survey phase of the mission. We will discuss the spectrometer
 in the next section. However, half the effective area of the spectrometer was
 utilised for a deep survey, observing at 90$^{\circ}$ to the main survey and
 giving exposure times significantly greater than the main sky survey but over
 a restricted region of the sky.

The scientific highlights of the ROSAT WFC and EUVE have been well documented
in the scientific literature as well as by \citet{bowyer00} and \citet{barstow03}. The principal outcome
of the sky surveys was the detection of several hundred sources of EUV
radiation. The WFC was launched in 1990 and carried out it survey between June
1990 and January 1991. A first release of source detections, the Bright Source
Catalogue, included 383 objects \citep{pounds93}. This number was increased to
479 in a later release, the 2 RE catalogue \citep{pye95}, with improvements in
attitude reconstruction, background screening and source detection methods.
EUVE was launched in June 1992 and, consequently, carried out its sky survey
approximately 2 years after the WFC.  A preliminary bright source list was
published by \citet{malina94} and followed by the first EUVE catalogue of 410
sources \citep{bowyer94} comprising 287 objects from the all-sky survey and 35
from the deep survey. A further 88 objects were drawn from the bright source
list identified through a variety of special processing techniques. Following
the main survey, during spectroscopic follow-up observations, the sky survey
telescopes continued to observe field in the so-called right angle
programme. The second EUVE source catalogue \citep{bowyer96} included objects
from this partial, deep survey of the sky as well as the re-processed all-sky
and deep surveys yielding a total of 734 sources.

Despite the wide range of sources that emit in the EUV, the 
vast majority of objects
detected fall in to two categories, hot white dwarfs and stellar coronal
sources. These are seen in roughly equal numbers, but with about 20\% more
coronal sources. In
contrast, there are only relatively small numbers of cataclysmic variables and
OB stars. It is also evident that there are windows in the ISM through which
it is possible to see out of the Galaxy, through the detection of about 2
extragalactic sources. It is not surprising that most of these are located at
high galactic latitude where the ISM column densities are lowest. 

The major survey results concern the white dwarf and coronal populations. The
photometric luminosities of the hottest white dwarfs are seen to be suppressed
when compared with pure H model atmospheres, implying that their atmospheres
contain significant quantities of heavier elements, identified from follow-up
UV spectroscopy as principally, C, N, O, Si, Fe and Ni (see
\citealt{barstow03}). It should be mentioned that the ability to detect strong
EUV continua from white dwarfs with effective temperatures as low as
$\sim$2$\times$10$^{4}$ K is a consequence of the low opacity of H in the EUV
which allows the emergence of EUV fluxes from deeper, hotter layers.
 The coronal sources tend to lie much closer to the solar
system, mostly within 50-60 pc, indicative of their lower luminosity, while
white dwarf detections extend as far as 200 pc in significant numbers. An
important subgroup of objects revealed by the surveys are a number of hot
white dwarfs in relatively close binary systems, where, at visible wavelengths, the low
luminosity white dwarf is hidden in the glare of its much brighter
companion. In the EUV, the situation is reversed due to the high EUV
luminosity of white dwarfs compared to coronal sources. These binaries are
analogous to the Sirius system, which is only resolved into two components
because of its relatively close proximity to Earth.

For completeness, it necessary to mention a third sky survey with the ALEXIS
satellite \citep{priedhorsky88}. ALEXIS made use of normal incidence mirrors
with multi-layer coatings, which, in principle, provide greater effective area
than grazing incidence optics for the same aperture. However, the multilayer
reflectivity has a narrow response spanning a few tens of \AA, which can limit
the overall photon flux, particularly for broad-band continuum sources such as
white dwarfs. Unfortunately, after launch one of the solar panels deployed
prematurely. Although the instruments eventually became operational, limited
attitude control and reconstruction made it difficult to analyse and
understand the data collected. More than 18 steady EUV sources were detected
but only one that was not already discovered by the WFC or EUVE. However, one
of the ALEXIS science goals was detection of bright transients. Five were
reported during the mission lifetime of which 3 were associated with known
cataclysmic variables.

\section{EUV spectroscopy}

Photometric all-sky surveys are the fundamental sources of important information regarding
the general properties of groups of objects in the EUV source
population. However, when considering individual objects in detail, the amount
of information that can be extracted from a few photometric data points is
somewhat limited. Therefore, the ability to obtain spectroscopic follow-up
observations is an important next step in the exploitation of any
observational window. The inclusion of a spectrograph on the EUVE mission was
an important addition, since it provided immediate access to spectroscopic
follow-up data on completion of the all-sky survey. The instrument made use of
the deep survey telescope placing three reflection gratings in the optical
path, each occupying about 1/6 of the collecting area, covering a total
wavelength range from 70 to 760 \AA~ in three segments: SW, 70-190 \AA; MW,
140-380 \AA; LW, 280-760 \AA. The gratings provided moderate resolving power
($\sim$ 200) across the band, with typical average values of 0.5 \AA, 1.0~\AA~
and 2.0 \AA~ for SW, MW and LW respectively.

With an effective area ranging from $\sim$ 0.2 to 2 cm$^{2}$, the EUVE
spectrographs were restricted mainly to observing the brightest EUV
sources. Even then, long exposure times were typically required from several
thousand to several tens of thousands of seconds. Nevertheless, many sources
were observed and some important and powerful diagnostic measurements
obtained. There is insufficient space in this review to give a full picture of
the work carried out, but we show a few examples of EUVE spectra and some of
the key highlights. The types of object observed divide simply into continuum
and line emission sources. The former are mostly white dwarfs but include a
number of cataclysmic variables and the B stars $\beta$ and $\epsilon$
CMa. The emission line sources are all stellar coronae.
\begin{landscape}
\begin{figure}
\begin{center}
\includegraphics*[scale=0.8,angle=270, trim=0cm 2cm 10cm 1cm,clip=true]{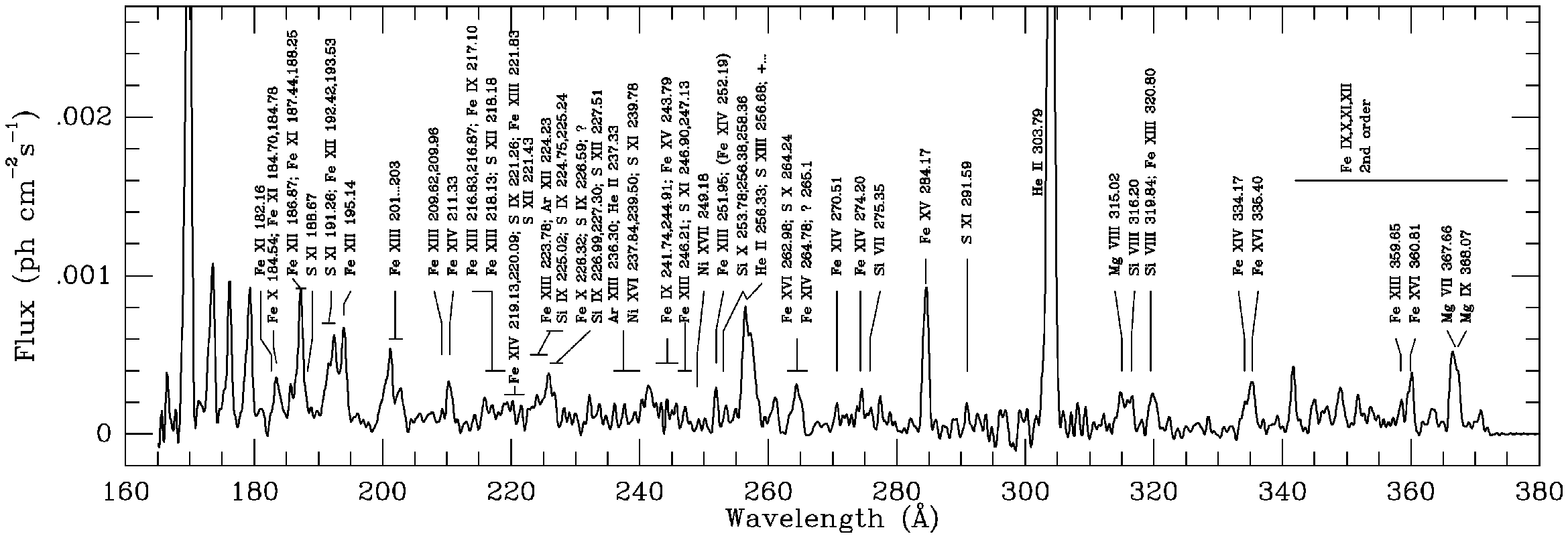}
\caption{\label{figure6}Flux calibrated MW EUV spectrum of Procyon. Individual lines and unresolved groups used in the analysis   of \citet{drake95}  are labelled, together with  a number of other  features not used.  The line identifier  '2nd'   refers to shorter wavelength lines seen in second order.}
\end{center}
\end{figure} 
\end{landscape}

Figure \ref{figure6} shows a section of one of the best cool star spectra
obtained by EUVE, of Procyon, illustrating the range of elements and
ionisation stages that can be observed \citep{drake95}. In a spectrum like
this there will be lines that are sensitive to the plasma temperature and
others that are density sensitive diagnostics. Combining this information
allows a determination of the physical properties of the coronal gas,
including the emission measure distribution and elemental abundances. 

\begin{figure}
\begin{center}
\includegraphics*[scale=1.2,angle=-0]{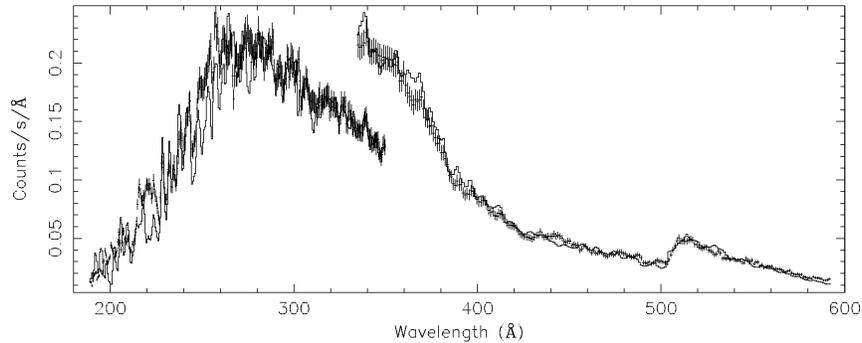}
\end{center}
\caption{\label{figure7} EUVE count spectrum of G191-B2B covering the wavelength range 180-600\AA. The data points (error bars) are compared with the predictions of a non-LTE model atmosphere calculation including the effects of interstellar absorption (H I, He I and He I column densities are 2.1 $\times$ 10$^{18}$, 1.8 $\times$ 10$^{17}$ and 7.9 $\times$ 10$^{17}$ cm$^{-2}$ respectively). The discontinuity near 320\AA~ arises from differing spectrometer effective areas for which these data are not corrected. }
\end{figure} 

In the continuum sources such as white dwarfs, spectral features are typically
seen in absorption. In the hot, dense atmospheres there can be a very large
number of high excitation features that cannot be resolved and appear as
blends. A good example is the H-rich DA white dwarf G191-B2B (Figure
\ref{figure7}). The short wavelength flux is strongly suppressed by the
opacity of heavy elements in the photosphere and there are many broad
absorption features, which are blends of many individual lines. Initially, for
stars like G191-B2B, the agreement between the data and stellar model
atmospheres was very poor. It was realised \citep{lanz96} that it was
necessary to include many more atomic transitions in the models to take
account of the true opacity in the stellar atmospheres. However, even though
the general shape of the spectra was reproduced, there was detailed
disagreement when modelling the individual features. In addition, to model the
shorter wavelength flux level, below 200 \AA, required development of
stratified atmospheres where the abundance of Fe in particular was allowed to
vary with depth \citep{barstow98}.

 \begin{figure}

\begin{center}
\includegraphics*[scale=2,angle=-0]{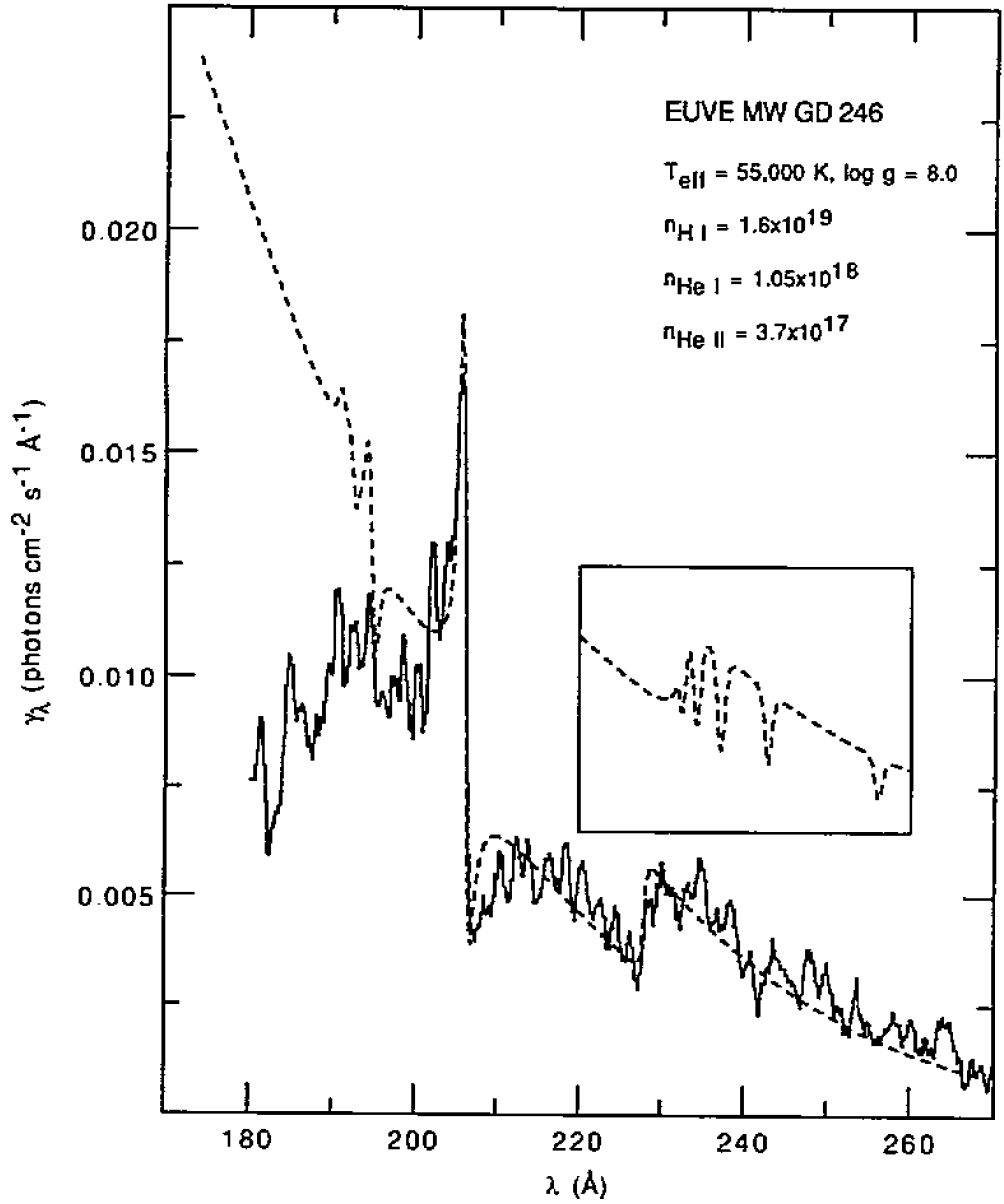}
\end{center}
\caption{\label{figure8}Observation of the DA white dwarf GD246 showing the region of the spectrum between 180 and 270 \AA~compared to a pure H atmosphere model including the interstellar opacity from H and He, showing the He I autoionisation feature at 206 \AA~and the He II absorption edge at 228 \AA~(from \citealt{vennes93}.}
\end{figure}

As outlined in the introduction, the EUV spectral range is unique in
containing the He I and He II resonance lines. At the resolution of EUVE the
individual lines in the series are not generally resolved but are evident in a
step function absorption ``edge'' generated by the blending of the lines near
the series limit. This can be seen in Figure \ref{figure7}, which show the He
I interstellar absorption edge at 504 \AA. In this example, the signature of
the He II edge at 228 \AA~and a He I autoionisation feature at 206 \AA~ are
blended with the photospheric absorption lines and hard to detect. However,
when the white dwarf atmosphere has fewer metals, these features can be seen
very clearly and provide powerful diagnostics on the physical state of the
local ISM (Figure \ref{figure8}). In particular, the fractional ionisation of
H and He can be determined and mapped across the local ISM. Interestingly,
within the observational errors, the H and He ionisation fractions appear to
be quite uniform (~23\% and 35\% respectively) in space
\citep{barstow97}. This has been interpreted as ionisation of the local
interstellar cloud surrounding the solar system by a past supernova explosion
followed by a period of recombination \citep{jenk13}. The current ionisation
fractions indicate an elapsed time of 2.1-3.4 million years since the onset
of recombination.

\section{High resolution spectroscopy with J-PEX}

The EUVE spectrometer made a number of very important advances in our
understanding of compact objects, coronal sources and the interstellar medium
through observations of several dozen of the brightest sources in the EUV
sky. Nevertheless, the limited effective area and spectral resolution
available placed considerable restrictions on the physical information that
could be extracted from the data. Particular issues arise in the presence of
large numbers of closely spaced spectral features that were unresolved. In
addition, limited dynamical information could be obtained from EUVE data,
since typical gas or spatial velocities were much smaller than the resolving
power. Therefore, it was inevitable that the science drivers following EUVE
would emphasise much improved spectral resolution and telescope collecting
area.
\begin{figure}
\begin{center}
\includegraphics*[width=13cm,angle=-0]{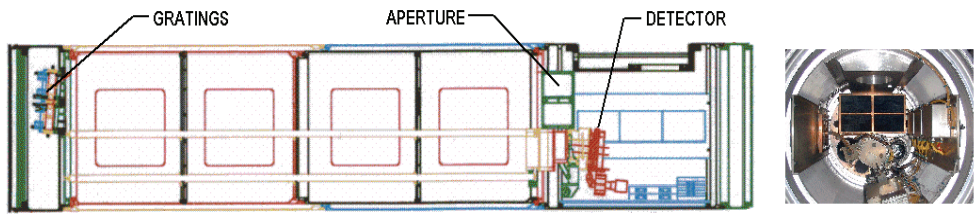}
\end{center}
\caption{\label{figure9}The J-PEX high-resolution spectrometer payload, designed and
  constructed for launch by a Terrier boosted Black Brant IX sounding rocket. The J-PEX spectrometer has made two successful flights, observing the isolated DA white dwarf G191-B2B and the DA+dM binary Feige 24, with the aim of examining their photospheric composition and searching for the presence of photospheric and interstellar He II. With a spectral resolution of 4000-6000 (depending on the flight configuration) it is possible to achieve the goal of developing the instrument, to resolve the individual lines of the He II Lyman series and separate them from other features from heavier elements. }
\end{figure} 

It is well understood that the grazing incidence optical systems used by the
WFC and EUVE are relatively inefficient light collectors, with large
collecting areas leading to large masses for the optics. The ALEXIS mission
pointed the way forward by producing lightweight telescopes based on
multilayer-coated normal incidence mirrors. However, these only provided
narrow band photometry.  The Joint astrophysical Plasmadynamic Experiment
(J-PEX) is a sounding rocket-borne spectrometer, constructed by the Naval
Research Laboratory, the University of Leicester and the Mullard Space Science
Laboratory, delivering both high throughput and spectroscopy. The instrument is
a slitless design, which employs a figured spherical grating in a Wadsworth
mount. This provides focusing and dispersion in a single optical element,
keeping reflectivity losses to a minimum.  As a result of practical upper
limits on the grating size at the time the payload was constructed, four
individual identical grating segments were utilised to maximise the geometric
collecting area. The design of the J PEX payload is shown in Figure
\ref{figure9}. The EUV light from the star enters a collimator, which
minimises the EUV background flux into the spectrometer, and strikes the
grating at an angle of incidence of 4.85 deg. The grating is coated with a
multilayer designed for high efficiency in the band 220-245 \AA. The
diffracted radiation is focused onto an MCP detector mounted on the grating
optical axis, a configuration, which minimises aberrations. Diffracted and
scattered EUV/FUV background is trapped by baffles (not shown) within the
spectrometer, while residual background reaching the detector is attenuated by
an aluminium filter. The instrument has been described in detail by
\citep{cruddace02} and \citep{barstow05}.

\begin{figure}

\begin{center}
\includegraphics*[scale=0.6,angle=270, trim=0cm 4cm 4cm 2cm,clip=true]{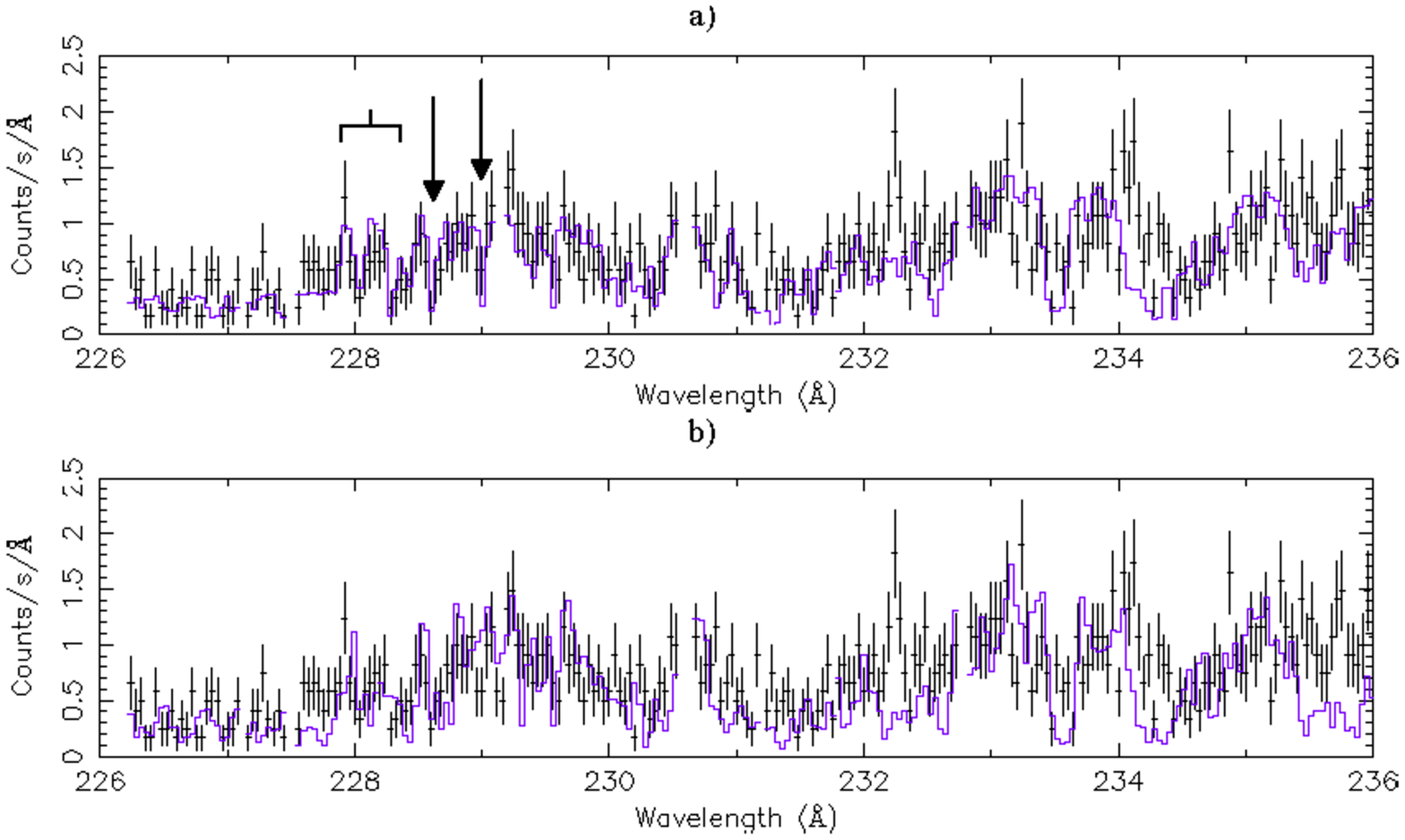}
\end{center}
\caption{\label{figure10}Comparison of the best-fit spectral models for each
  atmosphere type (histograms) with the 226 \AA~to 236 \AA~region of the
  J-PEX spectrum of G191-B2B (error bars). a) Homogeneous mixture of all
  elements, b) Self-consistent radiative levitation/diffusion PRO2 model. In
  a) the bracket identifies the converging He II Lyman series lines and the
  series limit, the left arrow marks HeII 228.54 \AA~and the right arrow a
  blended feature of O III and He II at 229.0 \AA.}
\end{figure} 

 \begin{figure}

\begin{center}
\includegraphics*[scale=0.6,angle=270, trim=0cm 4cm 4cm 2cm,clip=true]{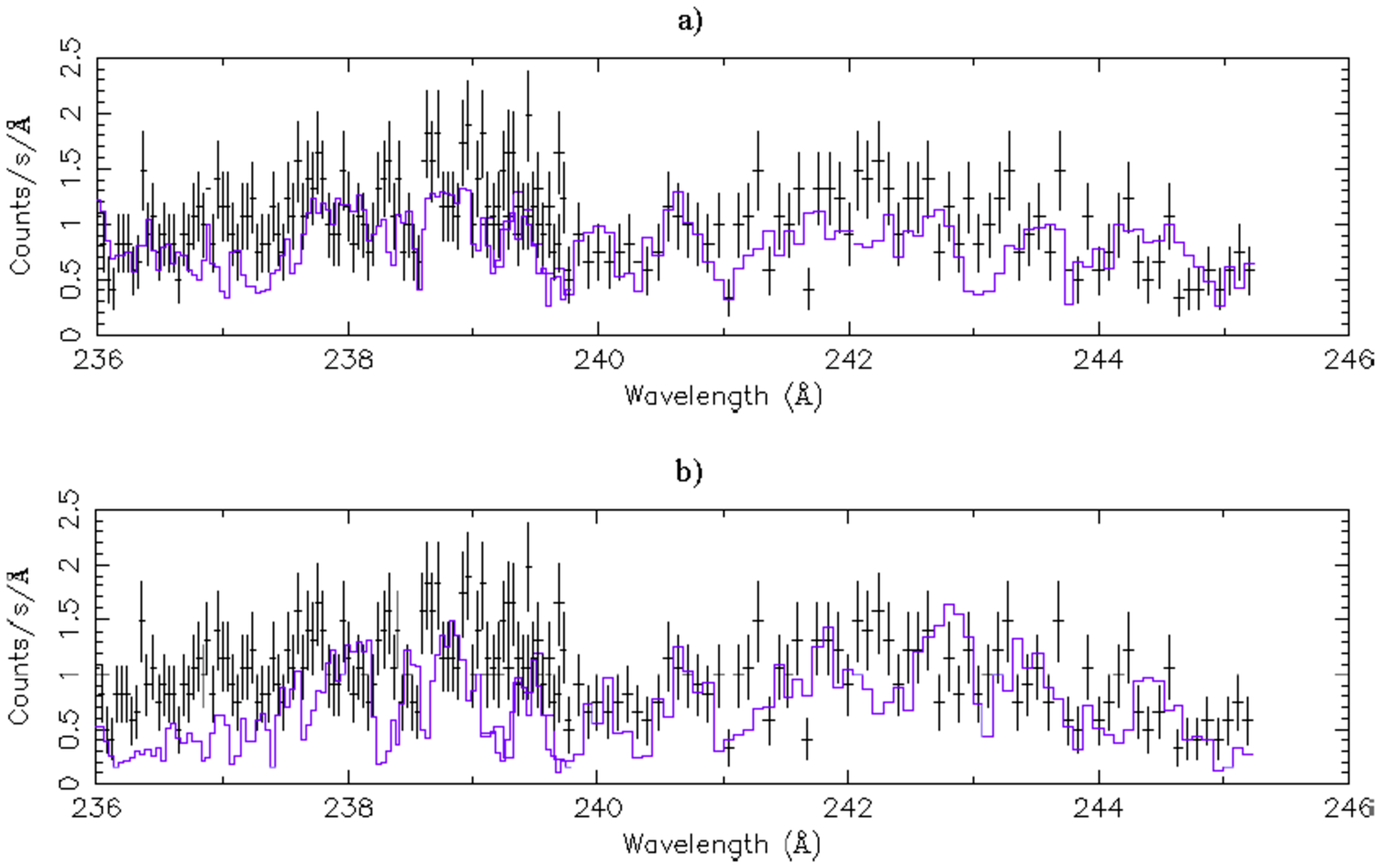}
\end{center}
\caption{\label{figure11} Comparison of the best-fit spectral models for each
  atmosphere type (histograms) with the 236 \AA~ to 246 \AA~ region of the J-PEX spectrum of G191-B2B (error bars). From the top a) Homogeneous mixture of all elements, b) Self-consistent radiative levitation/diffusion PRO2 model.}
\end{figure} 

Figures \ref{figure10} and \ref{figure11} show the J-PEX
spectrum of G191-B2B. \citet{cruddace02} provided an initial analysis of this
data and \citet{barstow05} carried out more detailed follow-up work. We
summarise the main results of \citet{barstow05} here who studied both
homogeneous and stratified mixtures of photospheric material. In contrast to
earlier work, based on EUVE spectra \citet{barstow98}, the J-PEX data are best
matched by the homogeneous models. This is illustrated in Figures
\ref{figure10} and \ref{figure11}, which compare the best fit between the two
types of stellar model atmospheres. Particularly good agreement is obtained in
the $\sim$ 228 \AA~ to 230 \AA~ wavelength range. Apart from the HeII Lyman
series, no strong absorption lines are predicted by the models in this
range. We can clearly identify $\lambda$ 228.54 \AA~ (left hand arrow in
Fig. \ref{figure10}), and blends of He lines down to the series limit (marked
by the bracket in Fig. \ref{figure10}). The apparent emission feature at 227.9
\AA~ is an artifact of the effect of He II line series opacity on the stellar
emission, which was predicted by modelling before the J-PEX flight. The
absorption line at 229.0 \AA~coincides with predicted He II and O III features
and is probably a blend of both (right hand arrow in Fig. \ref{figure10}). 

All EUV studies of G191-B2B have required an interstellar component (or
components) of HeII opacity to explain the observed spectra. At the resolution
of EUVE and in the presence of absorption from many other species, this
material could not be directly detected. However, the J-PEX data clearly
reveal the interstellar HeII Lyman series lines, and further they imply that
there is a contribution from photospheric He, although the latter is only
formally an upper limit, as the predicted strength of the 243 \AA \ line is similar to the noise in the spectrum (see Fig \ref{figure11}) and zero photospheric He is not a significantly different solution.

  \begin{figure}

\begin{center}
\includegraphics*[scale=0.8,angle=0, trim=3cm 3cm 0cm 8cm,clip=true]{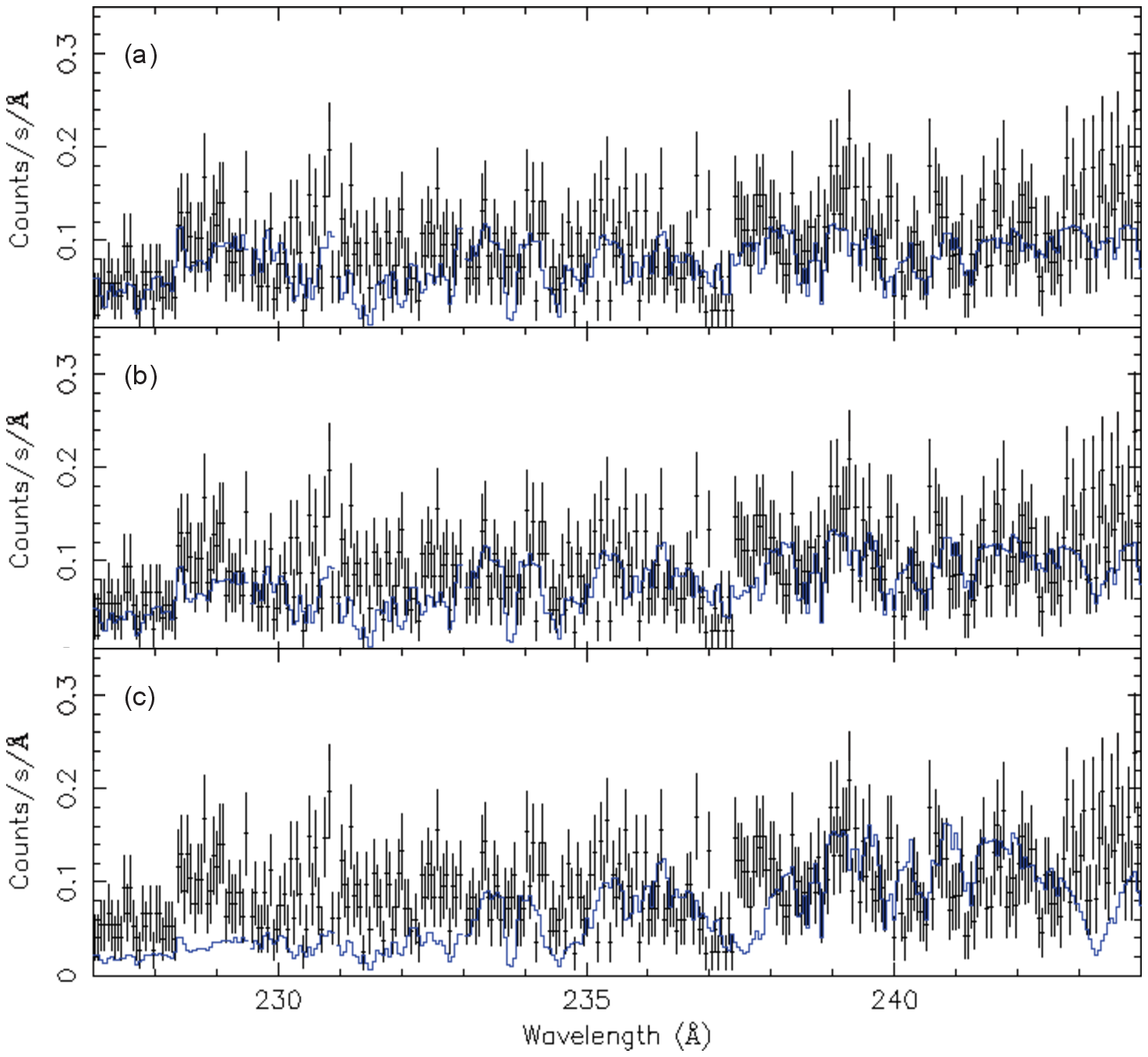}
\end{center}
\caption{\label{figure12} J-PEX high resolution spectrum of Feige 24 (black error bars) compared to a composite photospheric and ISM model spectrum (blue histogram). (a) Best fit model with H-layer mass of 1.2 $\times$ 10$^{-13}$ M$_{\odot}$, (b) H-layer mass = 3 $\times$ 10$^{-14}$ M$_{\odot}$, and c) H-layer mass = 10$^{-14}$ M$_{\odot}$.}
\end{figure} 

The DA+dM binary Feige 24 was observed as a comparison with G191-B2B. The two
white dwarfs have very similar temperatures and compositions, as measured by
UV spectroscopy. However, Feige 24 is a member of a close, 4.25d period,
binary system with its dM companion and will have passed through a common
envelope phase of evolution, which may have affected the structure of the
white dwarf. If this had happened, significantly reducing the mass of the H
envelope, this would be revealed through detection of photospheric He in the
EUV. Figure \ref{figure12} shows the spectrum recorded by J-PEX compared to
stellar model atmospheres with varying H envelopes. The very surprising result
of this work is that there is no evidence for the existence of a thin H-layer
or any photospheric He. This is illustrated in Figure \ref{figure12}(a), where
the H-layer mass has converged to the upper limit of the grid (1.2
$\times$10$^{-13}$ M$_{\odot}$), a level at which no He is detectable. Our
sensitivity to the H-layer mass is illustrated in Figures \ref{figure12}(b)
and \ref{figure12}(c), which show the effect of decreasing the H-layer mass in
the model to 3 $\times$ 10$^{-14}$ and 10$^{-14}$ M$_{\odot}$, respectively.  The strong
constraints on the amount of He present are provided by the absence of
detectable He II lines, particularly at 237.3 \AA~and 243 \AA, and the
predicted suppression of the continuum flux below $\sim$ 233 \AA~ for thinner
H-layers.

\section{Potential future missions: APEX and SIRIUS}

The J-PEX spectrograph has demonstrated the power of high-resolution
spectroscopy. Its high effective area provided a capability that could obtain
data of reasonable signal-to-noise in the few hundred seconds exposure
available in a sounding rocket flight. Nevertheless, the number of targets
that are bright enough for such observations is limited. Furthermore, to
observe a reasonable sample of objects would require an unrealistic number of
rocket flights. Hence, the J-PEX technology and the science it enables can
only be fully exploited if it is flown on a long duration satellite
platform. The J-PEX gratings were tuned to cover the wavelength range 220-260
\AA, addressing the science goals of studying the He II Lyman line series and
heavy elements in white dwarf spectra. The overall spectral coverage is
limited by the multilayer coating, where there is a tradeoff between
reflectivity and wavelength coverage. Maximising reflectivity reduces the
bandpass. J-PEX was an optimised solution to match the required wavelength
range for the science goals. Further important white dwarf and interstellar
medium science could be obtained over a wider wavelength range. In addition,
extending to shorter wavelengths would enable crucial observations of coronal
sources and cataclysmic variable stars, which were too faint for J-PEX. 

\begin{figure}

\begin{center}
\includegraphics*[width=13cm,angle=-0]{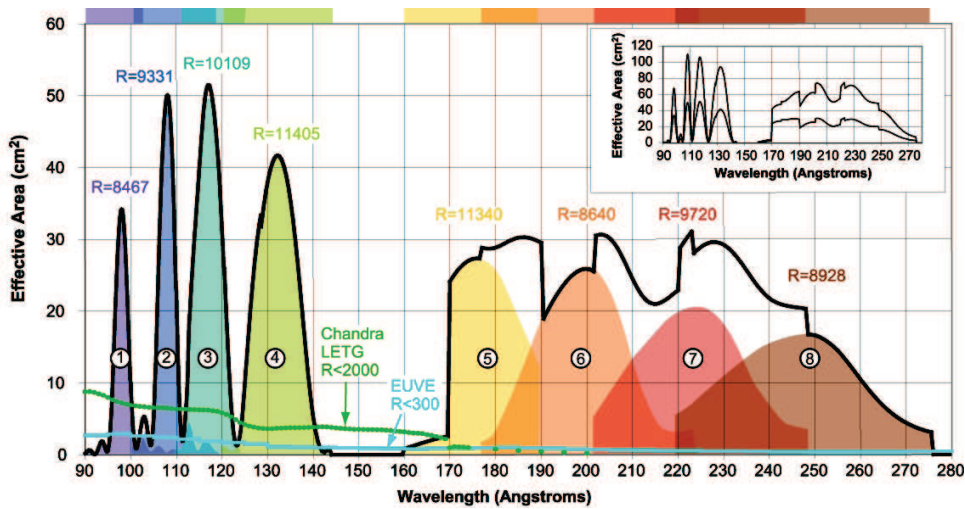}
\end{center}
\caption{\label{figure13} Baseline effective area of SAGE (inset: Baseline and Best-case effective area sums). The black curve is the cumulative effective area if the individual spectra are coadded.}
\end{figure}  

As a sounding rocket payload, J-PEX is a relatively low cost instrument with
proven technology at a high level of technological readiness. No further
technical development is necessary for translation to a satellite
mission. Therefore, it is also an excellent prospect for low cost satellite
development if coupled with an appropriate platform and launch
opportunity. The telescope has a 2 m long focal length enclosed in a $\sim$56
cm diameter cylindrical housing, which is small compared to typical large
space observatories such as Chandra, XMM-Newton or HST. There are
two main approaches to a satellite version of the spectrograph that have been
investigated to date. One is to
provide broad wavelength coverage by multiplexing several J-PEX style
telescopes, each tuned to a different waveband, on a single platform. Such a
package fits comfortably in to the usual physical and cost envelope of a small
to medium mission opportunity (depending upon the agency). A recent example
was the development of the SAGE concept in response to the recent ESA medium
mission opportunities. This utilises eight telescopes of 3 m focal length,
giving improved spectral resolution over J-PEX (see
\citealt{barstow09}). Figure \ref{figure13} shows the effective areas of each
telescope as a function of wavelength for a nominal payload that can study
stellar coronae, white dwarfs and the local interstellar medium compared to EUVE and the Chandra LETG. The dramatic increase in the performance is evident. While the LETG has a good spectral resolution it is not sufficient for the study of the dynamics of stellar coronae, where it is necessary to resolve velocities of a few tens of km/s.

\begin{figure}

\begin{center}
\includegraphics*[width=13cm,angle=-0]{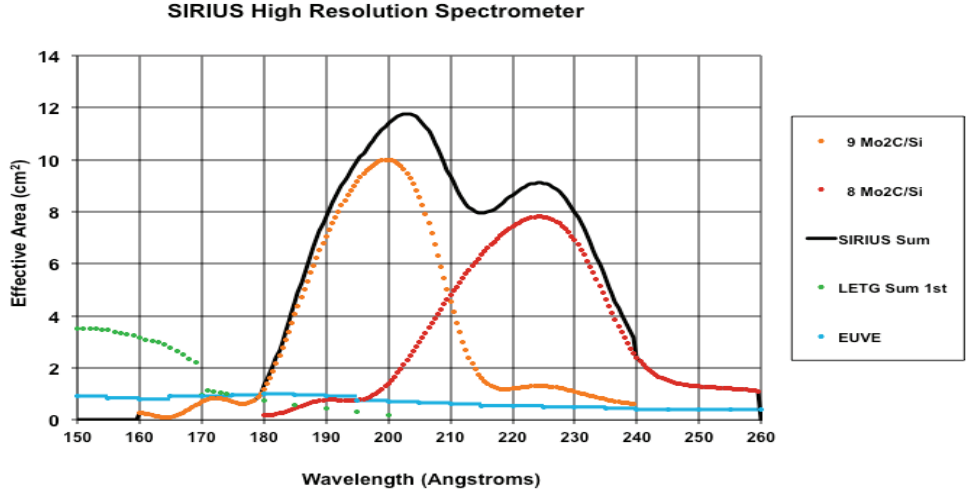}
\end{center}
\caption{\label{figure14} The effective area of SIRIUS, showing the individual wavebands, with Mo2C/Si coatings (orange and red), and the sum (black). Chandra LETG (green) and EUVE (blue) are shown for comparison.}
\end{figure} 

An alternative approach is make use of the modular nature of
the original J-PEX design, which had four identical gratings to deliver the
total required collecting area because it was not technically feasible to
produce a large enough monolithic grating at the time. In J-PEX the grating
multilayers were tuned to deliver the same wavelength coverage. The four
spectra were imaged individually and then coadded during data analysis to
achieve the full signal-to-noise available. However, with the longer exposures
available from a satellite platform, the four gratings could, in principle, be
tuned to different wavelength ranges to give broad coverage. This approach was
applied in the SIRIUS proposal in response to the 2012 ESA small (S) mission
call \citep{barstow12}. The four gratings were divided into two pairs, each
pair tuned to cover the wavelength range 17-260 \AA~(Figure
\ref{figure14}). Although this yields reduced effective area and spectral
coverage when compare to SAGE, much of the science is retained, allowing the
major goals of coronal, white dwarf and ISM studies to be achieved. Studies of
cataclysmic variables and extragalactic sources requiring even shorter
wavelength capabilities are sacrificed along the some of the shorter
timescales of variability studies. However, the SIRIUS payload can be
accommodated on a low cost satellite platform lower the mission cost to $\sim$
50M Euros, a factor of 5-6 lower than a typical medium scale mission envelope.

\begin{figure}[ht]
\begin{minipage}[b]{0.25\linewidth}
\centering
\includegraphics[scale=0.8,angle=0, trim=0cm 0cm 1cm 0cm,clip=true]{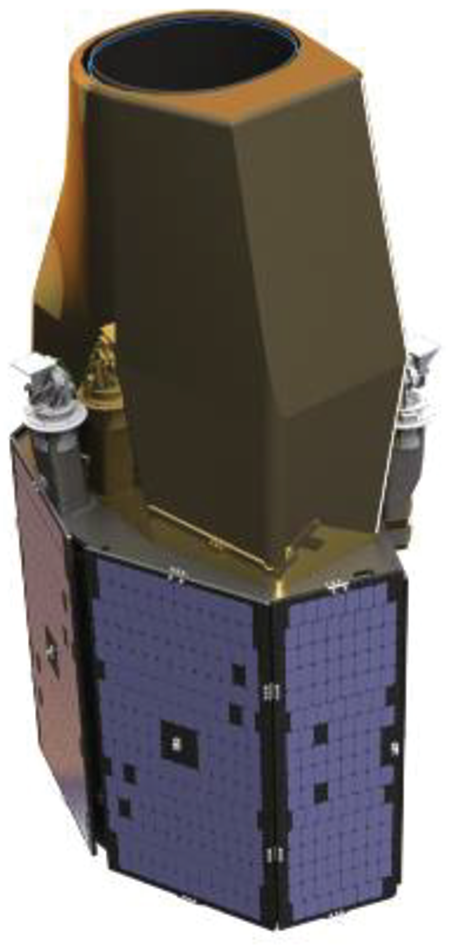}

\end{minipage}
\hspace{5.0cm}
\begin{minipage}[b]{0.25\linewidth}
\centering
\includegraphics[scale=0.8,angle=0, trim=0.5cm 0cm 0cm 0cm,clip=true]{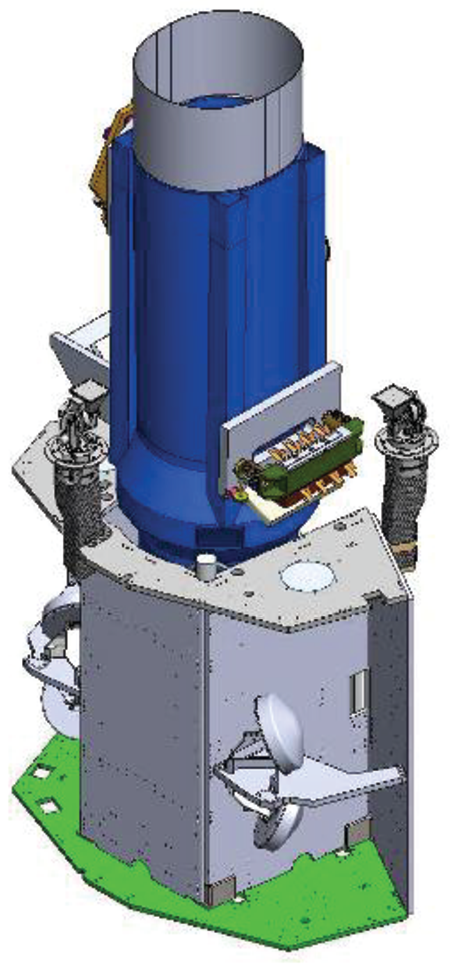}
\end{minipage}

\caption{\label{figure15}(left) The fully assembled SIRIUS satellite showing the spectrograph installed on the SSTL 300 spacecraft. (right) Cutaway, showing the telescope tube and spacecraft structure.}
\end{figure} 

The ability to provide a high resolution, high throughput
 EUV spectrometer as part of a low-cost mission arises not just from the
 extraordinary capability of the normal incidence grating technology but also
 from the availability of high performance, low cost spacecraft solutions such
 as the SSTL 300 platform that underpinned the SIRIUS proposal. The spacecraft
 consists of the payload (SIRIUS instrument) mounted in the aperture of the
 SSTL 300 spacecraft (Figure \ref{figure15}). The platform design draws on
 extensive heritage from previous missions, using a standard electrical
 architecture for subsystems and interfaces.

\section{Conclusion}

Although early assumptions about the opacity of the ISM consigned the EUV to
be a wavelength range where few if any objects could be studied outside the
solar system, observations at EUV wavelengths have been shown to be of great
important to astrophysics over the past $\sim$40 years development of
instrumentation to exploit this window in the electromagnetic spectrum. Sky
surveys have revealed more than 700 sources of EUV radiation. However,
interstellar opacity remains a dominant factor, as that vast majority of
sources are galactic in nature. Therefore, to a great extent EUV astronomy
will always be a niche for stellar astronomy and studies of the ISM. This
presents a significant hurdle in developing the next generation
high-resolution spectroscopy needed to further advance these subjects. An
extremely capable, observatory-class telescope such as SAGE can be
accommodated well within a medium class mission package. However, such a
mission is often competing with other proposals that address the current
priority areas of research such as cosmological studies or observations of
exoplanets. Therefore, it is likely that the most viable future opportunity
for EUV astronomy will be in the small mission arena, with a single telescope
flying on a low cost platform, such as the SIRIUS concept.



\end{document}